\documentclass[12pt,a4paper]{article} 
\usepackage{amssymb}
\usepackage{amsfonts}

\newtheorem{theorem}{Theorem}[section]

\newcommand{\ba}{\begin{array}}
\newcommand{\ea}{\end{array}}
\newcommand{\pa}{\partial}

\newcommand{\no}{\nonumber}

\newcommand{\be}{\begin{equation}}
\newcommand{\ee}{\end{equation}}
\newcommand{\bea}{\begin{eqnarray}}
\newcommand{\eea}{\end{eqnarray}}
\newcommand{\beaa}{\begin{eqnarray*}}
\newcommand{\eeaa}{\end{eqnarray*}}

\begin{document}

\title{The Gould-Hopper Polynomials  in the Novikov-Veselov equation}
\author{
 Jen-Hsu Chang \\Department of Computer Science and Information Engineering, \\
 National Defense University, \\
 Tauyuan, Taiwan }

\maketitle
\begin{abstract}
We use the Gould-Hopper (GH) polynomials to investigate the
Novikov-Veselov (NV) equation. The root dynamics of the
$\sigma$-flow in the NV equation is studied using the GH
polynomials and then the Lax pair is found. In particulr, when
$N=3,4,5$, one can get the Gold-fish model. The smooth rational
solutions of the NV equation are also constructed via the extended
Moutard transformation and the GH polynomials. The asymptotic
behavior is discussed and then the smooth rational solution of the
Liouville equation is obtained.

\end{abstract}



\newpage

\section{Introduction}
The Novikov-Veselov equation \cite{ba,gm,nv,vn}  is defined by
($U$ and $t$ is real) \bea
 U_t &=& \pa_z^3 U+ {\bar{\pa}}_z^3 U- 3\pa_z(VU)-3 {\bar{\pa}}_z
 (\bar{V}U ), \label{NV} \\
  {\bar{\pa}}_z V&=&\pa_z U. \no
 \eea
 When $z=\bar{z}=x$, we get the famous KdV equation ($U=V=\bar{V}$)
 \[ U_t=2U_{xxx}-12UU_x. \]
  The equation (\ref{NV}) can be represented as the form of Manakov's triad \cite{ma}
 \[H_t=[H, A]-BH, \]
 where $H$ is the two-dimensional Schrodinger operator
 \[H= \pa_z \bar{\pa_z} +U\]
 and
 \[A= \pa_z^3 - V\pa_z+{\bar{\pa}}_z^3-\bar{V}{\bar{\pa}}_z,
 \quad B= V_z+\bar{V}_{\bar{z}}.\]
 It is equivalent to the linear representation
 \bea H \psi=0, \quad \pa_t \psi=A \psi. \label{rep} \eea
 We see that the Novikov-Veselov equation (\ref{NV}) preserves a
 class of the purely potential self-adjoint operators $H$. Here
 the pure potential means $H$ has no external electric and
 magnetic fields. The periodic inverse spectral problem for the
 two-dimensional Schrodinger operator $H$ was investigated in
 terms of the Riemann surfaces with some group of involutions and
 the
 corresponding Prym $\Theta$-functions \cite{dk, gn, kr,mi,
 no,sh}. On the other hand, it is  known that the Novikov-Veselov hierarchy is a
 special reduction of the two-component BKP hierarchy \cite{wz,kt}(and references
 therein). In \cite{wz}, the authors showed that the Drinfeld-Sokolov
 hierarchy of D-type is a reduction of the  two-component BKP
 hierarchy using two different types of pseudo-differential
 operators, which is different from the Shiota's point of view \cite{sh}. Finally, it is worthwhile to notice that the Novikov-Veselov
 equation (\ref{NV})
 is a special reduction of the Davey-Stewartson equation
  \cite{ko, kg}.
 \\ \indent Let $H \psi= H \omega=0.$ Then via the Moutard transformation
 \cite{an, mo, ni}
 \beaa U(z, \bar{z}) &\longrightarrow & \hat{U}(z, \bar{z})=U(z, \bar{z})
 +2\pa \bar{\pa} \ln [i \int (\psi\pa \omega-\omega\pa \psi)dz-(\psi\bar{\pa} \omega-\omega\bar{\pa}
 \psi)d \bar{z}] \\
 \psi &\longrightarrow & \theta=\frac{i}{\omega} \int (\psi\pa \omega-\omega\pa \psi)dz-(\psi\bar{\pa} \omega-\omega\bar{\pa}
 \psi)d \bar{z},
 \eeaa
one can construct a new Schrodinger operator $\hat{H}= \pa_z
\bar{\pa_z} +\hat{U}$ and  $\hat {H}  \frac{1}{\theta}=0.$ The
extended Moutard transformation was established such that
\cite{hh, ms} \beaa \hat{U}(t,z,\bar{z}) &=& U(t,z,\bar{z})+2\pa
\bar{\pa} \ln i W(\psi, \omega),  \\
\hat{V} (t,z,\bar{z}) &=& V(t,z,\bar{z})+2 \pa \pa \ln i W (\psi,
\omega) \eeaa will also satisfy the Novikov-Veselov equation. Here
the skew product $W$ is defined by
 \bea W (\psi, \omega) &=& \int (\psi\pa
\omega-\omega\pa \psi)dz-(\psi\bar{\pa} \omega-\omega\bar{\pa}
 \psi)d \bar{z}+[\psi\pa^3 \omega-\omega\pa^3
 \psi+\omega\bar{\pa}^3 -\psi\bar{\pa}^3
 \omega   \no \\
&+&
2(\pa^2\psi\pa\omega-\pa\psi\pa^2\omega)-2(\bar{\pa}^2\psi\bar{\pa}\omega-\bar{\pa}\psi\bar{\pa}^2\omega)
 +3 V(\psi\pa \omega-\omega\pa \psi) \no \\
 &-&3\bar{V}(\psi\bar{\pa} \omega
 -\omega\bar{\pa}
 \psi)]dt.  \label{ext}
 \eea
In particular, we can use $U=V=0$ as the seed solution. Then $H=
\pa\bar{\pa}$. Let us consider the holomorphic function $P(z,t)$
and satisfy  \be \frac{\pa P}{\pa t}=\frac{\pa^3 P}{\pa z^3}
\label{lin}. \ee Then we have
\begin{theorem} \cite {tt} \\
Let $\mathcal{P}_1(t,z)$ and $\mathcal{P}_2(t,z)$  be different
holomorphic functions of
 $z$ and satisfy (\ref{lin}). One defines $\omega_1=\mathcal{P}_1+\bar{\mathcal{P}_1} $ and
$\omega_2=\mathcal{P}_2+\bar{\mathcal{P}_2} $. Then \bea
U(t,z,\bar{z}) &=& 2\pa \bar{\pa} \ln i W(\mathcal{P}_1 ,
\mathcal{P}_2), \label {u}\\
 V (t,z,\bar{z}) &=& 2 \pa \pa \ln i W (\mathcal{P}_1 ,
\mathcal{P}_2) \label{v} \eea is a solution of Novikov-Veselov
equation. Here the skew product $W$  is \bea
W (\mathcal{P}_1 , \mathcal{P}_2) &=& \mathcal{P}_1\bar{\mathcal{P}_2}-\mathcal{P}_2\bar{\mathcal{P}_1}+ \int [(\mathcal{P}_1'\mathcal{P}_2-\mathcal{P}_1\mathcal{P}_2')dz+(\bar{\mathcal{P}_1}\bar{\mathcal{P}_2}'-\bar{\mathcal{P}_1}'\bar{\mathcal{P}_2})d\bar{z} ] \no \\
&+& \int
[\mathcal{P}_1^{'''}\mathcal{P}_2-\mathcal{P}_1\mathcal{P}_2^{'''}+2
(\mathcal{P}_1'\mathcal{P}_2^{''}-\mathcal{P}_1^{''}\mathcal{P}_2')+\bar{\mathcal{P}_1}\bar{\mathcal{P}_2}^{'''}-\bar{\mathcal{P}_1}^{'''}\bar{\mathcal{P}_2}
\no \\
&+& 2(\bar{\mathcal{P}_1}^{''}\bar{\mathcal{P}_2}'
-\bar{\mathcal{P}_1}'\bar{\mathcal{P}_2}^{''})]dt
 .  \label{mou}
 \eea
 Moreover, if   $\mathcal{P}_1(t,z)$ and $\mathcal{P}_2(t,z)$     are polynomials,
then the solution is rational in $z, \bar{z}, t$.
\end{theorem}

\indent In \cite{bd,bf,df,gr}, the rational solutions and line
solitons of the Novikov-Veselov equation (\ref{NV}) were
constructed by the d-bar dressing method. To get these solutions,
the scattering data has to be delta-type and the reality of $U$
also puts some extra constraint on them. In these cases, the
$W$-function in  (\ref{u}) and (\ref{v}) can be expressed as a
determinant of some matrix. \\
\indent To study the dispersion relation (\ref{lin}), the authors
in \cite{tt} introduced the $\sigma$-flows for polynimial-type
solutions \be \mathbb{P}_N(t,z)=z^N +\sigma_1 z^{N-1}+\sigma_2
z^{N-2}+\cdots+\sigma_{N-1}z+\sigma_N. \label{si} \ee Then the
flow (\ref{lin}) generates a linear flow \bea
\dot{\sigma}_k=(N-K+3)(N-k+2)(N-k+1) \sigma_{k-3}, \quad k=1,2,3
\cdots N.\label{sig} \eea It can be seen that $\sigma_1, \sigma_2$
are conserved quantities. Indeed, $\sigma_1, \sigma_2, \cdots,
\sigma_N$ are the elementary symmetric polynomials in the roots
$q_1, q_2, \cdots, q_N$ of $\mathbb{P}_N (z)$: \bea
\sigma_1(\vec{q})&=&-\sum_{i=1}^N q_i, \qquad
\sigma_2(\vec{q})=\sum_{i< j} q_i q_j , \no \\
 \sigma_3(\vec{q})&=&-\sum_{i< j<k} q_i q_j q_k, \cdots,
\sigma_N(\vec{q})=(-1)^N q_1 q_2 \cdots q_N. \label{roo} \eea The
integrable (even linear) evolution of $\vec{\sigma}=(\sigma_1,
\sigma_2, \cdots, \sigma_N)$ induces a dynamical system on the
symmetric product $S^N C$ of the complex roots . We call
such a dynamical system on $S^N C$ a $\sigma$-system. \\
\indent From (\ref{mou}), we see that given two solutions
$\mathcal{P}_1(t,z)$ and $\mathcal{P}_2(t,z)$, one obtains a
solution of the Novikov-Veselov solution by a substitution of
$e^{i \lambda_1}\mathcal{P}_1(t,z)$ and $ e^{i
\lambda_2}\mathcal{P}_2(t,z)$ into (\ref{mou}). The  $\lambda_1$
and $\lambda_2$ defined here are real-valued constants. Therefore,
to each pair of holomorphic solutions of (\ref{lin}), we can get
an $(S^1 \times S^1)$-family of solutions to the
Novikiv-Veselov equation \cite{tt}. \\

\indent The paper is organized as follows. In section 2, we
describe the Gould-Hopper polynomials using the generating
function and establish the recursive relation.
In section 3, one studies the root dynamics of $\sigma$-flow and
the Lax pair is constructed. Also, the asymptotic behavior is
discussed.
In section 4, the smooth rational solutions are found using the
Gould-Hopper polynomials and their asymptotic behavior is
investigated.
Section 5 is devoted to the concluding remarks.

\section{Gould-Hopper Polynomials}
In this section, we introduce the Gould-Hopper polynomials and use
them to get the solutions of (\ref{lin}). To investigate the
polynomial-type solutions of (\ref{lin}), inspired by the work in
\cite{pg}, one  utilizes the Gould-Hopper polynomials \cite{gh}.
The generating function of the Gould-Hopper polynomials $P_N(t,z)$
is
\[ e^{\lambda z+ \lambda^3 t}=\sum_{N=0}^{\infty}P_N(t,z)
\frac{\lambda^N}{N!}. \] Indeed, the Gould-Hopper polynomials
$P_N(t,z)$ has the operator representation
\[P_N(t,z)=e^{t\pa_z^3}
z^N=[1+t\pa_z^3+\frac{t^2\pa_z^6}{2!}+\frac{t^3\pa_z^9}{3!}+\frac{t^4\pa_z^{12}}{4!}+\cdots]
z^N.\] Then one gets \be e^{t \pa_z^3} (e^{\lambda z})=e^{\lambda
z+ \lambda^3 t} \label{op}. \ee We remark that in general the
Gould-Hopper polynomials are
defined by $P^{(m)}_N(t,z)=e^{t\pa_z^m} z^N$.  Here we take $m=3$. \\
\indent One notices that the Gould-Hopper polynomials $P_N(t,z)$
are characterized by (\ref{lin}) and $P_N(0,z)=z^N$. For example,
\bea P_0 &=&1, \quad P_1=z, \quad P_2=z^2,
\quad P_3=z^3+6t, \quad  P_4=z^4+24tz, \no \\
P_5 &=&  z^5+60tz^2,  \quad P_6=z^6+120 tz^3+ 360t^2 \no \\
P_7&=& z^7+210tz^4+2520 z, \quad P_8=z^8+336tz^5+10080t^2z^2  \no \\
P_9&=& z^9+504 tz^6+30240 t^2z^3+60480t^3 \no  \\
P_{10}&=&z^{10}+720tz^7+75600t^2z^4+ 604800t^3z  \label{go}
 \eea
Actually, we have \bea P_N(t,z) &=& N! \sum_{k=0}^{[N/3]}
\frac{t^kz^{N-3k}}{k!(N-3k)!}  \no \\
\frac{d  P_N(t,z)}{dz}&=& N  P_{N-1} (t,z) \label{di} \eea From
the operation calculus, one has
\[ (z+3t\pa_z^2)P_{N-1}(t,z)=P_{N}(t,z), \quad  N \geq 1.\]
Hence we yield the recursive relation \be
P_{N}(t,z)=zP_{N-1}(t,z)+3t (N-1)(N-2) P_{N-3}(t,z).\label{cur}
\ee \indent We can see  that if we consider the equation
(\ref{lin}) with the initial data of analytical function
\[ P(0,z)=\sum_{N=0}^{\infty} \alpha_N z^N,\]
then the formal solution is \be
P(t,z)=e^{t\pa_z^3}\sum_{N=0}^{\infty} \alpha_N
z^N=\sum_{N=0}^{\infty} \alpha_N P_N(t,z).\label{sol} \ee The
 successive operation of the operator $(z+3t\pa_z^2)$ on the
 solution (\ref{sol}) can help us construct more solutions of
 (\ref{lin}). For example, if $P(0,z)= \sin z$, then we have,
 \beaa e^{t\pa_z^3} \sin z&=&e^{t\pa_z^3}
\sum_{N=0}^{\infty} \frac{(-1)^{N}}{(2N+1)!} z^{2N+1}  =
\sum_{N=0}^{\infty}\frac{(-1)^{N}}{(2N+1)!}P_{2N+1}(t,z)
\\&=& \sin(z-t). \eeaa
 The last equation uses (\ref{op}). Hence
\[(z+3t\pa_z^2)^N \sin(z-t),\quad  N=0, 1,2,3, 4, \cdots\]
are also solutions of (\ref{lin}). \\
\indent {\bf Remark:} Let's define
\[\varphi(\lambda)=e^{\lambda z+\lambda^3t}-e^{-\lambda z-\lambda^3
t}=2 sinh (\lambda z+\lambda^3 t).\] Then $\varphi(\lambda)$
satisfies
\[\varphi(\lambda)_{zz}=\lambda^2 \varphi(\lambda), \quad
\varphi(\lambda)_t= \varphi(\lambda)_{zzz}. \] On expanding
\[\varphi(\lambda)=\sum_{i=0}^{\infty}\phi_i \lambda^{2i+1}, \]
one has
\[\phi_{0,zz}=0, \quad \phi_{i+1,zz}=\phi_i, \quad \phi_{i,t}=\phi_{i,zzz},
\quad i\geq 0 .\] Actually,
\[ \phi_i= \sum_{k=0}^{[\frac{2i+1}{3}]} \frac{1}{k ! (2i+1-3k) !}
z^{2i+1-3k} t^k, \quad i\geq 0.\] It is known that  $\phi_i$ can
be used to construct the Wronskian solutions of the KdV equation.
The details can be found in \cite{my}.
\section{Root Dynamics of $\sigma$-flows}
In this section, one uses the Gould-Hopper polynomials to study
the root dynamics of the $\sigma$-flows (\ref{sig}). \\
\indent  Let's write (\ref{si})  as
\[\mathbb{P}_N(t,z)=(z-q_1(t))(z-q_2(t))\cdots (z-q_N(t)).\]
Then from the equation (\ref{lin}), one gets the root dynamics \be
\dot{q_j}=-6 \sum_{m<n,\quad j\neq m,n}^N
\frac{1}{(q_j-q_m)(q_j-q_n).} \label{main} \ee For example, when
N=3, we have \beaa
\dot{q_1}&=&-6 \frac{1}{(q_1-q_2)(q_1-q_3)} \\
\dot{q_2}&=&-6 \frac{1}{(q_2-q_1)(q_2-q_3)} \\
\dot{q_3}&=&-6 \frac{1}{(q_3-q_1)(q_3-q_2)} \eeaa For N=4, we get
\beaa
\dot{q_1}&=&-6[\frac{1}{(q_1-q_2)(q_1-q_3)}+\frac{1}{(q_1-q_3)(q_1-q_4)}+\frac{1}{(q_1-q_2)(q_1-q_4)}] \\
\dot{q_2}&=&-6[\frac{1}{(q_2-q_1)(q_2-q_3)}+\frac{1}{(q_2-q_3)(q_2-q_4)}+\frac{1}{(q_2-q_1)(q_2-q_4)}] \\
\dot{q_3}&=&-6[\frac{1}{(q_3-q_1)(q_3-q_2)}+\frac{1}{(q_3-q_1)(q_3-q_4)}+\frac{1}{(q_3-q_2)(q_3-q_4)}] \\
\dot{q_4}&=&-6[\frac{1}{(q_4-q_2)(q_4-q_3)}+\frac{1}{(q_4-q_1)(q_4-q_2)}+\frac{1}{(q_4-q_1)(q_1-q_3)}]
\eeaa We notice that since $\sigma_1$ and $\sigma_2$ are conserved
quantities, one knows that
\[ \sum_{i=1}^N q_i, \qquad \sum_{i=1}^N q_i^2 \]
are conserved densities of (\ref{main}). \\
\indent Now, we can  investigate the properties of the root
dynamics (\ref{main}) by the Gould-Hopper polynomials:
\begin{itemize}
\item Initial Value Problem : The root dynamics of $\sigma$-flow
can be solved by \bea \mathbb{P}_N(t,z) &=&
(z-q_1(t))(z-q_2(t))\cdots (z-q_N(t))  \no \\ &=& P_N(t,z)+
C_1P_{N-1}(t,z)+\cdots+C_NP_0(t,z), \label{inv} \eea where the
constants $C_1, C_2, \cdots, C_{N-1}, C_N$ are determined by the
initial values of $q_1(0), q_2(0), \cdots , q_N(0)$, that is,
\beaa C_1 &=&-\sum_{i=1}^N q_i(0), \qquad
C_2=\sum_{i< j} q_i(0) q_j(0) , \no \\
 C_3 &=& -\sum_{i< j<k} q_i(0) q_j(0) q_k(0), \quad \cdots, \no \\
C_N &=&(-1)^N q_1(0) q_2(0) \cdots q_N(0). \eeaa  Therefore, it is
seen that the solutions $q_1(t), q_2(t), \cdots , q_N(t)$ can be
obtained algebraically.
\item Lax pair:\\
Firstly,  we study the root dynamics of the Gould-Hopper
polynomials, which correspond to the initial values $ q_1(0)=
q_2(0)= \cdots = q_N(0)=0 $. \\
\indent \quad  Let's define the $N \times N$ matrix by \be
X(t)=\left\{
\begin{array}{lll}
     a_{i,i+1}=1, & \hbox{if $i=1,2,3, \cdots, N$ ;} \no \\
    a_{i,i-2}=-3t(i-1)(i-2), & \hbox{if $i=3,4 \cdots, N-1 $;} \no \\
    0, & \hbox{otherwise.} \label{mat}
\end{array}
\right.  \ee Then from the recursive relation (\ref{cur}), one
knows that
\[P_N(t,z)=det(zI_N-X(t)).\]
For example, when $N=3$, \beaa
 X(t)=
 \left(\ba{ccc} 0 & 1 & 0 \\
                0 & 0 & 1 \\
                -6t & 0 & 0 \ea \right); \eeaa
N=4,\beaa
 X(t)=
 \left(\ba{cccc} 0 & 1 & 0 &0 \\
                0 & 0 & 1 & 0 \\
                -6t & 0 & 0 &1 \\
                0& -18t & 0 &0\ea \right); \eeaa
N=5,\beaa
 X(t)=
 \left(\ba{ccccc} 0 & 1 & 0 &0 &0\\
                0 & 0 & 1 & 0 &0\\
                -6t & 0 & 0 &1 &0 \\
                0& -18t & 0 &0 &1 \\
                 0& 0&-36t & 0 &0 \ea \right).\eeaa

We can write $X(t)$ as
\[X(t)=R(t)QR^{-1}(t), \]
where $Q=diag(q_1(t), q_2(t), \cdots, q_N(t)$ and \be
 R(t)=
 \left(\ba{ccccc} P_0(q_1,t)& P_0(q_2,t) & P_0(q_3,t) & \cdots &P_0(q_N,t)\\
                P_1(q_1,t)& P_1(q_2,t) & P_1(q_3,t) & \cdots &P_1(q_N,t)\\
                P_2(q_1,t)& P_2(q_2,t) & P_2(q_3,t) & \cdots & P_2(q_N,t) \\
                \vdots \\
                 P_N(q_1,t)& P_N(q_2,t) & P_N(q_3,t) & \cdots &P_N(q_N,t)\ea \right).\label{rt} \ee
For instance, when $N=3$, \beaa
 R(t)=
 \left(\ba{ccc} 1 & 1 & 1 \\
                q_1 & q_2& q_3 \\
                q_1^2 & q_2^2 & q_3^2 \ea \right); \eeaa
and N=4, \beaa
 R(t)=
 \left(\ba{cccc} 1 & 1 & 1 &1\\
                q_1 & q_2& q_3 & q_4\\
                q_1^2 & q_2^2 & q_3^2 & q_4^2 \\
                  q_1^3+6t & q_2^3+6t & q_3^3+6t & q_4^3+6t \ea \right); \eeaa
N=5, \beaa
 R(t)=
 \left(\ba{ccccc} 1 & 1 & 1 &1 &1  \\
                q_1 & q_2& q_3 & q_4 & q_5\\
                q_1^2 & q_2^2 & q_3^2 & q_4^2 & q_5^2\\
                  q_1^3+6t & q_2^3+6t & q_3^3+6t & q_4^3+6t & q_5^3 +6t \\
                 q_1^4+24t & q_2^4+24t & q_3^4+24t & q_4^4+24t & q_5^4 +24t \ea \right). \eeaa

From the initial value problem (\ref{inv}), we notice that the
polynomials $t^n$ can be replaced by the elementary symmetric
polynomials of the roots $q_1, q_2, \cdots, q_N $. Hence one has
$R(\vec{q})$. It can be seen that
\[\dot{X}(t)=RLR^{-1},\]
where
\[L=\dot{Q}+[M,Q], \quad M=R^{-1} \dot{R}.\]
For example, when N=3, \beaa
 L(t)=
 \left(\ba{ccc} \dot{q_1} & \dot{q_2}\frac{q_2-q_3}{q_3-q_1} & \dot{q_3}\frac{q_3-q_2}{q_2-q_1}\\
                 \dot{q_1}\frac{q_1-q_3}{q_3-q_2}& \dot{q_2}& \dot{q_3}\frac{q_3-q_1}{q_1-q_2} \\
                \dot{q_1}\frac{q_1-q_2}{q_2-q_3} & \dot{q_2}\frac{q_2-q_1}{q_1-q_3}& \dot{q_3} \ea \right);
                \eeaa
and N=4, using the Maple software, \beaa
 L(t)=
 \left(\ba{cccc} \dot{q_1} &- \frac{\dot{q_2}(q_2-q_3)(q_2-q_4)+6}{(q_1-q_3)(q_1-q_4)} &- \frac{\dot{q_3}
 (q_3-q_2)(q_3-q_4)+6}{(q_1-q_2)(q_1-q_4)} &- \frac{\dot{q_4}
 (q_4-q_2)(q_4-q_3)+6}{(q_1-q_2)(q_1-q_3)} \\- \frac{\dot{q_1}
 (q_1-q_3)(q_1-q_4)+6}{(q_2-q_3)(q_2-q_4)}& \dot{q_2}& -\frac{\dot{q_3}(q_3-q_1)(q_3-q_4)+6}{(q_2-q_4)(q_2-q_1)}
  &- \frac{\dot{q_4}(q_4-q_1)(q_4-q_3)+6}{(q_2-q_3)(q_2-q_1)} \\ -\frac{ \dot{q_1} (q_1-q_2)(q_1-q_4)+6}{(q_2-q_3)(q_4-q_3)} & -\frac{ \dot{q_2}
 (q_2-q_4)(q_2-q_1)+6}{(q_4-q_3)(q_1-q_3)}& \dot{q_3} & - \frac{\dot{q_4}
 (q_4-q_1)(q_4-q_2)+6}{(q_2-q_3)(q_1-q_3)}  \\- \frac{\dot{q_1}
 (q_1-q_2)(q_1-q_3)+6}{(q_4-q_3)(q_4-q_2)} & -\frac{ \dot{q_2}(q_2-q_3)(q_2-q_1)+6}{(q_4-q_3)(q_4-q_1)} & -\frac{ \dot{q_3}
 (q_3-q_1)(q_3-q_2)+6}{(q_4-q_1)(q_4-q_2)} &  \dot{q_4} \ea
 \right). \eeaa
Since \[\dot{X}(t)=\frac{d X(t)}{dt}=\left\{
\begin{array}{ll}
    a_{i,i-2}=-3(i-1)(i-2), & \hbox{if $i=3,4 \cdots, N-1 $;} \\
    0, & \hbox{otherwise,}  \\
\end{array}
\right. \] we know $\dot{X}(t)$ is a nilpotent matrix and hence
$L$ is a nilpotent one, too. So
\[ tr(L^r)=tr[\dot{X}(t)]^r=0, \quad r=1,2, 3, \dots, ..\]
Actually, a simple calculation yields
\[L^{[\frac{N}{2}]+1}=0, \quad N \geq 3,\]
where $[\alpha]$ means the integer part of $\alpha$. \indent Now,
\[\frac{d^2X(t)}{dt^2}=0\]
will imply the Lax equation \be \frac{dL(t)}{dt}=[L,M].
\label{lax} \ee For $N=3,4,5$, we see that, by the Maple software,
$q_i$ satisfies the following Goldfish model \cite{ca}, a limiting
case of the Ruijesenaars-Schneider system: \be \ddot{q}_i =2
\sum_{j \neq i} \frac{q_i q_j}{q_i-q_j}. \label{gold} \ee The
reason is that $P_i, i=3,4,5$ are linear in $t$-variable(see the
appendix). For $N=6$, we have from the diagonal terms of the Lax
equation (\ref{lax}) \[ \ddot{q}_i =2 \sum_{j \neq i}^6 \frac{q_i
q_j}{q_i-q_j} +\frac{\sum_{j=1}^6 \mbox{(some quadratic terms of
$\vec{q}$)}q_j+720} {\prod_{i\neq j}^6(q_i-q_j)}. \] \indent \quad
Secondly, we consider the general case. Let's define the 2D Appell
polynomials $\mathbb{G}_n(z,t)$ by means of the generating
function \cite{br}: \be A(\lambda) e^{\lambda z+ \lambda^3
t}=\sum_{n=0}^{\infty}\mathbb{G}_n (z,t) \frac{\lambda^n}{ n
!},\label{ap} \ee where
\[ A(\lambda) =\sum_0 ^N \frac{\mathbf{\Gamma}_k}{k!} \lambda^k,\]
$\mathbf{\Gamma}_k's$ being constants and $\mathbf{\Gamma}_0=1$.
Then one has the following  formula, noting that ${N \choose h}={N
\choose N-h}$,  \be \mathbb{G}_N = \sum_{h=0}^N {N \choose N-h}
\mathbf{\Gamma}_{N-h} P_h(z,t). \label{app} \ee It's easy to see
that the polynomials $\mathbb{G}_n(z,t)$ also satisfy the linear
equation (\ref{lin}). When comparing (\ref{inv}) with (\ref{app}),
we have \be \mathbf{\Gamma}_{N-h}=\frac{C_{N-h}}{{N \choose N-
h}}.\label{in} \ee \indent \quad Now, it's suitable to introduce
the coefficients of the Taylor expansion
\[ \frac{A'(\lambda)}{A(\lambda)}=\sum_{n=0}^{\infty} \alpha_n
\frac{\lambda^n}{n!}. \] It can be seen that the coefficients
$\alpha_n$ can be expressed by $\mathbf{\Gamma}_0,
\mathbf{\Gamma}_1, \cdots, \mathbf{\Gamma}_{n+1}$ (or the initial
values (\ref{in})). For examples, \beaa \alpha_0 &=&
\mathbf{\Gamma}_1, \quad \alpha_1= \mathbf{\Gamma}_2-
\mathbf{\Gamma}_1^2, \quad  \alpha_2= \mathbf{\Gamma}_3- 3
\mathbf{\Gamma}_1  \mathbf{\Gamma}_2+2  \mathbf{\Gamma}_1^3, \\
\alpha_3 &=&  \mathbf{\Gamma}_4+ 12  \mathbf{\Gamma}_1^2
\mathbf{\Gamma}_2-4
\mathbf{\Gamma}_1  \mathbf{\Gamma}_3-3 \mathbf{\Gamma}_2^2-6 \mathbf{\Gamma}_1^4, \\
&\vdots & \eeaa The recurrence relation for the 2D Appell
polynomial $\mathbb{G}_N(z,t)$ can be
written as follows \cite{br}: \bea \mathbb{G}_{0}(z,t) &=& 1   \no \\
\mathbb{G}_N(z,t) &=& (z+\alpha_0) \mathbb{G}_{N-1}(z,t)+ 3t
(N-1)(N-2)\mathbb{G}_{N-3}(z,t)  \no \\
&+&\sum_{k=0}^{N-2} {N-1 \choose k}
\alpha_{N-k-1}\mathbb{G}_{k}(z,t). \label{rec} \eea  A simple
calculation can yield \beaa \mathbb{G}_{0}(z,t) &=& 1,
\quad \mathbb{G}_{1}(z,t)=z+\alpha_0 , \quad \mathbb{G}_{2}(z,t)=(z+\alpha_0)^2+\alpha_1 \\
\mathbb{G}_{3}(z,t)&=&(z+\alpha_0)^3+3\alpha_1(z+\alpha_0)+
\alpha_2+ 6t \\
\mathbb{G}_{4}(z,t)&=&(z+\alpha_0)^4+6\alpha_1(z+\alpha_0)^2+
(4\alpha_2+24t)(z+\alpha_0)+ \alpha_3+3\alpha_1^2 \\
\mathbb{G}_{5}(z,t)&=&(z+\alpha_0)^5+10 \alpha_1(z+\alpha_0)^3+
(10\alpha_2+60t)(z+\alpha_0)^2 \\
&+&
(5\alpha_3+15\alpha_1^2)(z+\alpha_0)+60t\alpha_1+10\alpha_1\alpha_2+\alpha_4.
\eeaa

When $A(\lambda)=1$, this recursive relation becomes (\ref{cur}).
Hence the relation (\ref{rec}) is a generalization of (\ref{cur})
for arbitrary initial data.  The matrix corresponding to
(\ref{mat}) can be constructed as follows: \beaa X(t)=\left\{
\begin{array}{lllllllll}
     a_{i,i+1}=1, & \hbox{if $i=1,2,3, \cdots, N-1$ ;}  \\
     a_{i,i}=-\alpha_0, & \hbox{if $i=1,2,3, \cdots, N$ ;} \\
a_{i,i-1}=-{i-1 \choose i-2}\alpha_1, & \hbox{if $i=2,3, 4 ,\cdots, N$ ;} \\
a_{i,i-2}=-{i-1 \choose i-3}(6t+\alpha_2), & \hbox{if $i=3,4,5 , \cdots, N$ ;} \\
a_{i,i-3}=-{i-1 \choose i-4}\alpha_3, & \hbox{if $i=4, 5, 6, \cdots, N$ ;} \\
\vdots \\
a_{i,i-k}=-{i-1 \choose i-k-1}\alpha_k, & \hbox{if $i=k+1,k+2, k+3, \cdots, N$ ;} \\
\vdots \\
a_{N,1}=-\alpha_{N-1}
\end{array}
\right.  \eeaa
Similarly, one has
\[\mathbb{G}_{N}(z,t)=det(zI_N-X(t)).\]
For instance, when $N=5$, we get \beaa
 X(t)=
 \left(\ba{ccccc} -\alpha_0 & 1 & 0 &0 &0\\
                -\alpha_1 & -\alpha_0 & 1 & 0 &0\\
                -(6t+\alpha_2) & -2\alpha_1& -\alpha_0 &1 &0 \\
                -\alpha_3 & -(18t+3\alpha_2) & -3\alpha_1 &-\alpha_0 &1 \\
                 -\alpha_4& -4\alpha_3&-(36t +6\alpha_2)& -4\alpha_1 &-\alpha_0 \ea \right).\eeaa
Also, one can write $X(t)$ as
\[X(t)=R(t)QR^{-1}(t), \]
where $Q=diag(q_1(t), q_2(t), \cdots, q_N(t)$ and $R(t)$ is
defined as (\ref{rt}) with $P_m(q_i,t)$ being replaced by
$\mathbb{G}_{m}(q_i,t)$. Then one follows the previous procedures
and finally can get the Lax equation (\ref{lax}) for general case.
Therefore the root dynamics (\ref{main}) is Lax-integrable. But
the computations are more involved and one doesn't pursuit them
here.
\\
\indent \quad We notice here that for $N=3,4,5$ the root dynamics
of $\mathbb{G}_N$ also satisfies the Gold-fish model (\ref{gold}).
\item Asymptotic behavior \\
It is known that the  Gould-Hopper polynomial $P_N(t.z)$ has the
scaling property: \be P_N(t,z)=t^{\frac{N}{3}}
\hat{P}_N(\frac{z}{t^{1/3}}), \label{sca}\ee where
$\hat{P}_N(\eta)$ is the so-called Appell polynomials \cite{br}
(and references therein) in $\eta=\frac{z}{t^{1/3}}$. For example,
\beaa
P_8(t,z)&=& z^8+336tz^5+10080t^2z^2=t^{\frac{8}{3}}[\eta^8+336 \eta^5+10080 \eta^2]\\
&=&t^{\frac{8}{3}}\hat{P}_8(\eta). \eeaa Then the k-th zero
$\lambda_N^{(k)} $ of $\hat{P}_N(\eta)$ determines the dynamics of
the root $q_k$, i.e.,
\[q_k(t)=\lambda_N^{(k)}t^{1/3}.\]
Since $\hat{P}_N(\xi \lambda_N^{(k)})=0, \xi^3=1,$ one knows that
the roots $q_k$ are located on the circles in the plane with time
dependent radius or fixed at the origin. Finally, from the Initial
value Problem (\ref{inv}) and (\ref{sca}), we know that when
$t\rightarrow \infty$ and $z\rightarrow \infty$ such that
$|z|^3/t\rightarrow $ constant, $P_N(t,z)$ plays the dominant
role. Then one yields
\[q_k(t) \rightarrow \lambda_N^{(k)}t^{1/3}. \]
Consequently, the roots asymptotically  will follow diagonal
lines.
\end{itemize}
\section{Smooth Rational Solutions of Novikov-Vaselov equation}
In \cite{tt}, the  blow-up solution of the  Novikov-Vaselov
equation via the extended Moutard transformation (\ref{mou}) is
constructed. In this section, we establish  smooth rational
solutions for all time by the Gould -Hopper polynomials (\ref{go}). Some calculations below need the Maple software. \\
\begin{itemize}
\item Example 1 \\
Let $\mathcal{P}_1=z^2+z+1$ and $\mathcal{P}_2=-iz^2-2iz$. Then a
simple calculation can yield the imaginary part of $W$ in
(\ref{mou}) \be M(x,y,t)=(x^2+y^2)^2+ \frac{8}{3}x^3+
4xy^2+4x^2+4x+4t+100
 \label{ex0}
\ee From (\ref{ex0}), we can see that
\[ M(x,y,0) \approx (x^2+y^2)^2 \quad \mbox{near} \quad  r=\sqrt{x^2+y^2}=\infty.  \]
It can be verified that $M(x,y,0)$ is positive for all
$\mathbb{R}^2$.  Also,  $M(x,y,t)$ is positive for all
$\mathbb{R}^2$ at any fixed time  $t \geq 0$ . Then the solution
$U$ in (\ref{u}) of the Novikov-Veselov equation (\ref{NV}) is
\[ U=\frac{M_1}{M_2},\] where
\beaa M_1 &=& -12\,
(294+600\,{x}^{2}+588\,{y}^{2}+8\,{x}^{3}+888\,x+12\,t+3\,{
x}^{4}-6\,{x}^{2}{y}^{2}-2\,{x}^{3}{y}^{2}
\\
&+& 24\,{x}^{2}t-3\,{y}^{4}x+24
\,{y}^{2}t+36\,xt-3\,{y}^{4}+{x}^{5} ) \eeaa and \beaa M_2 &=&
\left( 3\,{x}^{4}+6\,{x}^{2}{y}
^{2}+3\,{y}^{4}+8\,{x}^{3}+12\,x{y}^{2}+12\,{x}^{2} \right. \\
&+& \left. 12\,x+12\,t+300 \right) ^{2} \eeaa At fixed time t ,
one knows $U$ decays like $\frac{1}{r^3}$ for $r \rightarrow
\infty$. Also, $U$ tends asymptotically  to zero at the rate
$\frac{1}{t}$ at any fixed point $(x,y)$ when $t \rightarrow
\infty$.

\item Example 2 \\
Let $\mathcal{P}_1=(z^3+6t)+2 i z$ and
$\mathcal{P}_2=-i(z^3+6t)+ z$. Then a simple calculation can yield
the imaginary part of $W$ in (\ref{mou}) \be f(x,y,t)=(x^2+y^2)^3+
4x^3y+ 8xy^3+2(x^2+y^2)+ 6t(2x^3-6xy^2-y)+36t^2+6000
 \label{ex1}
\ee From (\ref{ex1}), we can see that
\[ f(x,y,0) \approx (x^2+y^2)^3 \quad \mbox{near} \quad  r=\sqrt{x^2+y^2}=\infty.  \]
It can be verified that $f(x,y,0)$ is positive for all
$\mathbb{R}^2$. Also, letting (\ref{ex1}) be equal to zero, one
has \beaa
t &=& (1/2)xy^2+(1/12)y-(1/6)x^3 \\
&\pm &
(1/12)\sqrt{24x^2y^4-20xy^3-36x^4y^2-7y^2-20x^3y-4y^6-8x^2-23996},
\eeaa A simple calculation shows that the equation inside the
square root is negative for all $\mathbb{R}^2$. Hence $f(x,y,t)$
is positive for all $\mathbb{R}^2$ at any fixed time $ t\geq 0$.
Then the solution $U$ in (\ref{u}) of the Novikov-Veselov equation
(\ref{NV}) is
\[ U=\frac{F_1}{F_2},\] where \beaa F_1 &=& -\frac{1}{2}[24
\,{x}^{7}y+16\,{x}^{6}+24\,{x}^{5}{y}^{3}+ \left( 432\,ty+
216000+48\,{y}^{2} \right) {x}^{4} \\
&+&  \left( -24\,{y}^{5}-144\,t+48\,y
 \right) {x}^{3} + \left( 432000\,{y}^{2}+864\,t{y}^{3}-96\,{y}^{4}
 \right) {x}^{2} \\
 &+&
  \left( -24\,{y}^{7}-48\,{y}^{3}+432\,t{y}^{2}+
 \left( 432000+1728\,{t}^{2} \right) y \right) x+48000 \\
 &+& 432\,{y}^{5}t-
32\,{y}^{6}+216000\,{y}^{4}+252\,{t}^{2}]  \eeaa and \beaa F_2
&=&\left( {x}^{6}+3\,{x}^{4}{y
}^{2}+3\,{x}^{2}{y}^{4}+{y}^{6}+4\,{x}^{3}y+8\,x{y}^{3}+2\,{x}^{2}+2\,
{y}^{2} \right.\\
&+& \left. 12\,t{x}^{3}-36\,tx{y}^{2}-6\,ty+36\,{t}^{2}+6000
\right) ^{2} \eeaa At fixed time t , one knows $U$ decays like
$\frac{1}{r^4}$  for $r \rightarrow \infty$. Also, $U$ tends
asymptotically  to zero at the rate  $\frac{1}{t^2}$ at any fixed
point $(x,y)$ when $t \rightarrow \infty$.

\item Example 3\\
Let $\mathcal{P}_1=(z^4+24tz)+2 i z$ and
$\mathcal{P}_2=-i(z^4+24tz)+ z$. A tedious calculation shows that
the imaginary part of $W$ in (\ref{mou}) is
\[g(x,y,t)=(x^2+y^2)^4+6x^4y+12x^2y^3-(18/5)y^5+2x^2+2y^2+24t(x^2+y^2)(2x^3-6xy^2+24t)+100      \]
Similarly, $g(x,y,t)$ is positive for all $\mathbb{R}^2$ at fixed
time  $t \geq 0$.  Then the solution $U$ in (\ref{u}) can be
written as
\[ U=\frac{G_1}{G_2}, \]
where \beaa && G_1(x,y,t)
\\
&=&-\frac{1}{2}[240\,{y}^{11}-1680\,{x}^{2}{y}^{9}-1980\,{y}^{8}+
\left( 34560 \,tx-5280\,{x}^{4} \right) {y}^{7} \\
&+ & \left( 160000-2160\,{x}^{2}
 \right) {y}^{6}
 + \left( 480+218880\,t{x}^{3}-3360\,{x}^{6}+138240\,{t
}^{2} \right) {y}^{5} \\
&+& \left( 480000\,{x}^{2}-9000\,{x}^{4} \right) {y }^{4}  +
\left( -120000+1200\,{x}^{8}+57600\,t{x}^{5}  \right.\\
&+&  \left.\left( -1382400\, {t}^{2}-4800 \right) {x}^{2} \right)
{y}^{3} +  \left( 3600\,{x}^{6}- 5760000\,tx+480000\,{x}^{4}
\right) {y}^{2}  \\
&+&  \left( 1200\,{x}^{10}+ 57600\,{x}^{7}t+ \left(
2400+691200\,{t}^{2} \right) {x}^{4}+360000\,{ x}^{2} \right) y \\
&+& 20000+160000\,{x}^{6}+1920000\,t{x}^{3}+5760000\,{t}^
{2}+900\,{x}^{8} ]  \eeaa and \beaa
 && G_2(x,y,t) \\
 &=& \left(
5\,{x}^{8}+20\,{x}^{6}{y}^{2}+30\,{x}^{4}{y}
^{4}+20\,{x}^{2}{y}^{6}+5\,{y}^{8}+30\,{x}^{4}y+60\,{x}^{2}{y}^{3}-18
\,{y}^{5}+10\,{x}^{2} \right. \\
&+& \left. 10\,{y}^{2}+240\,t{x}^{5}-480\,t{x}^{3}{y}^{2}+
2880\,{x}^{2}{t}^{2}-720\,t{y}^{4}x+2880\,{y}^{2}{t}^{2}+500
\right) ^ {2} \eeaa In this case, at fixed time, one knows that
$U$ decays like $\frac{1}{r^5}$ for $r \rightarrow \infty$ ;
moreover, $U$ also tends asymptotically to zero at the rate
$\frac{1}{t^2}$ at fixed point $(x,y)\neq (0,0)$ when $t
\rightarrow \infty $. We notice here that at $(0,0)$,  $U$
approaches $\infty$  as $t \rightarrow \infty$.
\item Example 4\\
Let $\mathcal{P}_1=(z^5+60 tz^2)+2 i z$ and
$\mathcal{P}_2=-i(z^5+60tz^2)+ z$. Then the imaginary part of $W$
in (\ref{mou}) is \beaa h(x,y,t)&=&\left(  \left( {x}^{2}+{y}^{2}
\right) ^{5}-\frac{10}{3}\,{x}^{5}y+{ \frac
{20}{3}}\,{x}^{3}{y}^{3}+ \left( {x}^{2}+{y}^{2} \right)
 \left( 2+12\,{x}^{3}y-12\,x{y}^{3} \right) \right.\\
 &+& \left.120\,t \left( {x}^{2}+{y}
^{2} \right)  \left( {x}^{5}-3\,{y}^{4}x-2\,{x}^{3}{y}^{2} \right)
+ 3600\,{t}^{2} \left( {x}^{2}+{y}^{2} \right) ^{2} \right.\\
&+& \left.20\,t \left( 11\,{y} ^{3}+3\,{x}^{2}y \right)
+120\,{t}^{2}+1000 \right) \eeaa Similarly, $h(x,y,t)$ is positive
for all $\mathbb{R}^2$ at fixed time $t \geq 0$.  Then the
solution $U$ in (\ref{u}) is
\[U=\frac{H_1}{H_2} , \]
where \beaa  && H_1(x,y,t) \\
&=& -\frac{1}{2}[360\,x{y}^{13}+
\left( 360\,{x}^{3}+14400\,t \right) {y}^{11}- 144\,{y}^{10}+
\left(
14400\,{x}^{2}t-1800\,{x}^{5} \right) {y}^{9} \\
&+& \left( -720\,{x}^{2}+900000+108000\,{t}^{2} \right) {y}^{8} +
\left( 237600\,t{x}^{4}-777600\,{t}^{2}x-3600\,{x}^{7} \right)
{y}^{7}   \\
&+& \left( 3840\,{x}^{4}+ \left( 3600000+432000\,{t}^{2} \right)
{x}^{2}+ 83520\,tx \right) {y}^{6}+ \left( 165600\,t{x}^{6}+720\,x
\right.
 \\
 &-& \left.10368000\,{t }^{3}-1800\,{x}^{9}-345600\,{t}^{2}{x}^{3} \right)
{y}^{5}+ \left( - 1120\,{x}^{6}+ \left( 5400000 \right.\right.
\\ &+&  \left. \left.648000\,{t}^{2} \right) {x}^{4}   + 12000\,t {x}^{3}+
\left( -129600000\,t-15552000\,{t}^{3} \right) x-810000\,{t}^ {2}
\right) {y}^{4} \\
&+& \left( 360\,{x}^{11} - 50400\,{x}^{8}t+950400\,{x}^
{5}{t}^{2}-2160\,{x}^{3}+20736000\,{t}^{3}{x}^{2}+ \left(
-216000\,{t} ^{2} \right. \right.\\
&-&\left. \left. 1800000 \right) x -5760\,t \right) {y}^{3}+
\left( 2380\,{x}^{8}+
 \left( 3600000+432000\,{t}^{2} \right) {x}^{6}-25200\,t{x}^{5} \right.\\
 &+& \left.
 \left( -10368000\,{t}^{3}-86400000\,t \right) {x}^{3}+453600\,{x}^{2}
{t}^{2}+62208000\,{t}^{4}+518400000\,{t}^{2} \right) {y}^{2} \\
&+&
\left(360\,{x}^{13}+21600\,{x}^{10}t+518400\,{x}^{7}{t}^{2}+720\,{x}^{5}+
31104000\,{t}^{3}{x}^{4}+ \left( 1920000
\right.\right.\\
&+& \left.\left.230400\,{t}^{2} \right) {x}^{
3}+17280\,{x}^{2}t+1555200\,{t}^{3}+12960000\,t \right)
y+476\,{x}^{10 }  \\
&+& \left( 900000 +108000\,{t}^{2} \right)
{x}^{8}+25200\,{x}^{7}t+
 \left( 43200000\,t+5184000\,{t}^{3} \right) {x}^{5} \\
 &+& \left.226800\,{x}^{4}{t}^{2}+ \left( 62208000\,{t}^{4}  +518400000\,{t}^{2}\right) {x}^{2}+
72000+8640\,{t}^{2} \right)] \eeaa and \beaa && H_2(x,y,t) \\
&=&
\left( 3\,{x}^{10}+15\,{x}^{8}{y}^{2}+30\,{x}^{6
}{y}^{4}+30\,{y}^{6}{x}^{4}+15\,{y}^{8}{x}^{2}
 + 3\,{y}^{10}+26\,{x}^{5}
y+20\,{x}^{3}{y}^{3}\right.\\
&+& \left. 6\,{x}^{2}+6\,{y}^{2}-36\,x{y}^{5}+360\,{x}^{7}t-
1800\,t{x}^{3}{y}^{4}-360\,t{x}^{5}{y}^{2}-1080\,tx{y}^{6} \right.
\\
&+& \left.10800\,{x}^
{4}{t}^{2}+21600\,{t}^{2}{x}^{2}{y}^{2}+10800\,{t}^{2}{y}^{4}+660\,t{y
}^{3}+180\,t{x}^{2}y+360\,{t}^{2}+3000 \right) ^{2}  \eeaa In this
case, at fixed time, $U$ decays like $\frac{1}{r^6}$ for $ r
\rightarrow \infty$ ; moreover, we see that \be U \rightarrow
\frac{-240(x^2+y^2)}{[30(x^2+y^2)^2+1]^2}=\frac{-240
z\bar{z}}{[30z^2\bar{z}^2+1]^2} \quad as \quad t \rightarrow
\infty \label{as} \ee at fixed point $(x,y)$, which is a
stationary solution of (\ref{NV}) for
\[V= {\frac {3600\,{\bar{z}}^{4}{z}^{2}-120\,{\bar{z}}^{2}}{ \left( 30\,{z}^{2}{\bar{z}}^{2}+
1 \right) ^{2}}}. \]
\end{itemize} We notice that if we define
\[u(z,\bar{z})=\ln U \quad and \quad V(z,
\bar{z})=\frac{U_{zz}}{3U}, \] then the stationary equation of the
Novikov-Veselov equation  (\ref{NV}) will become the Tzitzeica
equation \cite{fe,mi2} \be u_{z\bar{z}}=e^u+ \epsilon e^{-2u}
\label{tz}, \ee where $\epsilon$ is an arbitrary constant. It can
be verified that (\ref{as}) satisfies the $\epsilon=0$ case, i.e.,
the Liouville equation, whose real solutions are given by \be \ln
\frac{-2\kappa | \frac{dS}{dz}|^2}{(1+\kappa |S|^2)^2},\label{s}
\ee  where $S(z)$ is a locally univalent meromorphic function in
some domain and $\kappa$ is a constant. For the solution
corresponding to (\ref{as}), one knows
that $S(z)=z^2$ and $\kappa=30$. \\
\indent For  general case, let's choose
\[\mathcal{P}_1=P_N(t,z)+2 i z \quad  and \quad
\mathcal{P}_2=-iP_N(t,z)+ z . \] We can expect the imaginary part
of $W$ in (\ref{mou}) is positive for all $\mathbb{R}^2$ at any
fixed time  if we choose an appropriate constant. And the solution
$U(x,y,t)$ at any fixed time decays like $\frac{1}{r^{N+1}}$ for
$N \geq 3$. For $N=1$, one refers to \cite{gr}. It could be
interesting to study the asymptotic solutions when $t \rightarrow
\infty$ and then one obtains the solutions (\ref{s}) of the
Liouville equation or the ones of the Tzitzeica equation
(\ref{tz}). One remarks here that in these examples, for each
potential $U(x,y,t)$ there exist infinitely many wave functions,
which can be constructed by the Pfaffian \cite {an, ni} of linear
combinations of the Gould-Hopper polynomials. Hence it could be
worthwhile to investigate the corresponding wave functions when $t
\rightarrow \infty$. But the computations are more involved and
the details will be published elsewhere.

\section{Concluding Remarks}
In this paper we have studied the Novikov-Veselov equation using
the Gould-Hopper polynomials. Firstly, one investigates the root
dynamics of the so-called $\sigma$-flows and gets the Lax pair;
moreover, one finds that when $N=3,4,5$, the root dynamics
satisfies the Gold-Fish  model. Although the Lax pair is
established, only two conserved densities are found. The reason is
that the Lax operator is nilpotent. Also, the asymptotic behavior
is studied. Secondly, we construct smooth rational solutions using
the Gould-Hopper polynomials and the skew product (\ref{mou});
besides, the asymptotic behaviors of the smooth rational solutions
are discussed.
\subsection*{Acknowledgments}
The author is grateful to Prof. Iskander Taimanov and Prof. Sergy
Tsarev for valuable discussions when they visited Taiwan. He also
thanks Prof. Wen-Xiu Ma for his suggestions. This work is
supported in part by the National Science Council of Taiwan under
Grant No. NSC 98-2115-M-606-001-MY2.
\newpage
\renewcommand{\theequation}{A.\arabic{equation}}
\section{Appendix}
\appendix
\setcounter{equation}0
\section{Gold-Fish Model}
For the Goldfish Model
\[ \ddot{q}_i =2 \sum_{j \neq i} \frac{\dot{q}_i \dot{q}_j}{\dot{q}_i-q_j}, \]
its initial value problem can be solved by the statement:  $
z=q_i(t),i=1,2, \cdots, N $ are the N roots of the equation
\cite{ca}
\[\sum_{i=1}^N \frac{\dot{q}_i(0)}{z-q_i(0)}=\frac{1}{t}.\]
It can be seen that it is a polynomial in $z$ with coefficients
linear in $t$.  Then the special choices of initial datum can get
the solutions of the root dynamics (\ref{main}) for the cases
$N=3,4,5$. To illustrate it, we take $N=3$ as an example. When
$N=3$, we have
\[ \frac{\dot{q}_1(0)}{z-q_1(0)}+ \frac{\dot{q}_2(0)}{z-q_2(0)}+\frac{\dot{q}_3(0)}{z-q_3(0)}=\frac{1}{t}.\]
After some calculations, one yields \bea
&& z^3-z^2[(q_1(0)+q_2(0)+q_3(0)]+z[q_1(0)q_2(0)+q_2(0)q_3(0)+q_1(0)q_3(0)]-q_1(0)q_2(0)q_3(0) \no \\
&&=
tz^2[\dot{q}_1(0)+\dot{q}_2(0)+\dot{q}_3(0)]-tz[\dot{q}_1(0)(q_2(0)+q_3(0))+\dot{q}_2(0)(q_1(0)+q_3(0))
\no \\ &&+\dot{q}_3(0)(q_1(0)+q_2(0))]+t
[\dot{q}_1(0)q_2(0)q_3(0)+\dot{q}_2(0)q_1(0)q_3(0)+\dot{q}_3(0)q_1(0)q_2(0)].
\label{equ}  \eea On the other hand, from (\ref{inv}), one knows
$q_1(t), q_2(t), q_3(t)$ are the roots of the polynomial \[
\mathbb{P}_3(z,t)=z^3+ 6t +C_1z^2+C_2 z+C_3 \] or \be
  z^3 +C_1z^2+C_2 z+C_3 =-6t. \label{po} \ee Comparing (\ref{equ}) with (\ref{po}),  we are able to get the following linear equations for $ \dot{q}_1(0),
\dot{q}_2(0), \dot{q}_3(0)$ : \bea &&
\dot{q}_1(0)+\dot{q}_2(0)+\dot{q}_3(0)=0
\no \\
&& \dot{q}_1(0)(q_2(0)+q_3(0))+\dot{q}_2(0)(q_1(0)+q_3(0))
+\dot{q}_3(0)(q_1(0)+q_2(0))=0 \no \\
&&
\dot{q}_1(0)q_2(0)q_3(0)+\dot{q}_2(0)q_1(0)q_3(0)+\dot{q}_3(0)q_1(0)q_2(0)=-6.
\label{ear} \eea So if the determinant of the matrix
\be \left(\ba{ccc} 1 & 1 & 1 \\
                q_2(0)+q_3(0)& q_1(0)+q_3(0) & q_1(0)+q_2(0) \\
                 q_2(0)q_3(0)& q_1(0)q_3(0) & q_1(0)q_2(0) \ea
                 \right) \label{mr}
                 \ee
is not equal to zero, then the initial velocities $ \dot{q}_1(0),
\dot{q}_2(0), \dot{q}_3(0)$ can be uniquely expressed by the
initial positions $q_1(0), q_2(0), q_3(0)$. For the cases $N=4,
5$, the linear equations (\ref{ear}) and the matrix (\ref{mr}) can
be obtained similarly.

\end{document}